\documentclass[
aps,%
final,%
notitlepage,%
oneside,%
onecolumn,%
nobibnotes,%
nofootinbib,%
superscriptaddress,%
noshowpacs,%
centertags]%
{revtex4}

\usepackage{amsmath,amssymb}
\usepackage[bookmarks=false,colorlinks=true]{hyperref}
\usepackage[usenames,dvipsnames]{xcolor}
\usepackage[usenames,dvipsnames]{xcolor}
\usepackage{latexsym}
\usepackage{bm}
\newcommand{\be}{\begin{equation}}
\newcommand{\ee}{\end{equation}}

\newcommand{\bea}{\begin{eqnarray}}
\newcommand{\eea}{\end{eqnarray}}

 \definecolor{armygreen}{rgb}{0.29, 0.33, 0.13}

 \include{rgb}

\begin{document}

\title{Lobachevsky geometry in  TTW and PW systems}
\author{Tigran Hakobyan}
\email{tigran.hakobyan@ysu.am}
\author{Armen Nersessian}
\email{arnerses@ysu.am}
\affiliation{Yerevan State University, 1 Alex Manoogian St., Yerevan, 0025, Armenia}
\affiliation{Tomsk Polytechnic University, Lenin Ave. 30, 634050 Tomsk, Russia}
\author{Hovhannes Shmavonyan}
\email{hovhannes.shmavonyan@ysumail.am}
\affiliation{Yerevan State University, 1 Alex Manoogian St., Yerevan, 0025, Armenia}

\begin{abstract}
We review the classical properties of  Tremblay-Turbiner-Winternitz and Post-Wintenitz systems
and their relation  with $N$-dimensional rational Calogero model with oscillator and Coulomb potentials,
paying special attention to their hidden symmetries. Then we show that combining the radial coordinate
and momentum in a single complex coordinate in proper way,
we get an elegant description for the hidden and dynamical symmetries in these systems
related with action-angle variables.
\end{abstract}
\maketitle

\section{Preliminaries}
\noindent
The rational  Calogero model and its generalizations, based on arbitrary Coxeter root systems,  are  highlighted  among  the non-trivial unbound
superintegrable systems.
Recall that the superintegrability of $N$-dimensional integrable system means that it possesses $2N-1$ functionally independent constants of motion.
This property was established for the classical  \cite{woj83} and quantum \cite{kuznetsov,gonera}
rational  Calogero model, which is described by the Hamiltonian \cite{calogero}
\be
H_0=
 \sum_{i=1}^N \frac{p_i^2}{2} + \sum_{i<j} \frac{g^2}{(x_i-x_j)^2}.
\label{calo}
\ee
Its generalization, associated
with an arbitrary finite Coxeter group, is defined by the Hamiltonian \cite{rev-olsh}
\be
\label{coxeter}
H_0=
 \sum_{i=1}^N \frac{p_i^2}{2} + \sum_{\alpha\in\mathcal{R}_+}
\frac{g^{2}_\alpha (\alpha\cdot\alpha)}{2(\alpha\cdot x)^2}.
\ee
Let us mention that the Coxeter group is described
 as a finite group  generated
 by a set of orthogonal reflections  across the hyperplanes $\alpha\cdot x=0$
 in the $N$-dimensional Euclidean space,
where the vectors $\alpha$ from the set $\mathcal{R}_+$  (called the system of positive roots)
 uniquely characterize the reflections.
The coupling constants $g_\alpha$ form a reflection-invariant discrete function.
The original Calogero potential in \eqref{calo} corresponds to the $A_{N-1}$ Coxeter system with
the positive roots, defined in terms of the standard basis by $\alpha_{ij}=e_i-e_j$
for $i<j$. The reflections become the coordinate
permutations in this particular case.

An important feature of all  the rational Calogero models is the dynamical conformal
symmetry   $so(1,2)\equiv sl(2,R)$, defined  by  the Hamiltonian $H_0$
together  with the   dilatation $D$ and  conformal boost $K$ generators,
\begin{equation} \{  H_0 , D\}=2 H_0  ,\qquad\{
H_0 , K\}=D,\qquad  \{ K, D\}= -2K. \label{ca} \end{equation}
This symmetry separates  the radial  and angular parts  as follows:
\be
{H}_0=\frac{p^2_r}{2}+\frac{{\cal I}(u)}{r^2},
\qquad  r\equiv \sqrt{2K},
\qquad
 p_r\equiv \frac{D}{\sqrt{2K}},
 \ee
 where
 \be \{p_r, r\}=1,\quad \{p_r, u^\alpha\}=\{r, u^\alpha\}=0,\quad \{u^\alpha,u^\beta\}=\omega^{\alpha\beta}(u).
\label{sC}
\ee

Hence, the whole information about the rational Calogero model (and, more generally, any conformal mechanics) is encoded in its "spherical part",
given by the Hamiltonian $\mathcal{I}(u)$  (which corresponds to the Casimir element of the conformal algebra).
The "angular Calogero model" given by the Hamiltonian $\mathcal{I}$ was studied  from the various viewpoints in \cite{sphCal,lny,flp}.
 Apart from its inherent interest, this system provides  the rational Calogero models  with an elegant explanation of  maximal superintegrability \cite{hlkn}.
 This property is retained for other root systems, and persists  in the presence of an additional oscillator potential
 (Calogero-oscillator system).
 Moreover,  by the separation of the radial and angular parts,
 we have recently established  the superintegrability for the rational Calogero model
 with Coulomb potential (Calogero-Coulomb system),
 as well as for the Calogero-oscillator and Calogero-Coulomb systems on
 sphere and hyperboloid.
 Similar statements are valid for the Calogero-like models associated with arbitrary root systems.
 In case of $A_N$ Calogero-Coulomb model, we have presented an explicit
 expression for the analog of Runge-Lenz vector \cite{runge}, and  revealed its   integrable generalizations
 for the two-center Coulomb  (two-center Calogero-Coulomb)
 and Stark (Calogero-Coulomb-Stark) potentials \cite{Calogero-Stark}.

The   Trembley-Turbiner-Wintenitz (TTW) system, invented a few years ago \cite{TTW},
is a particular case of the Calogero-oscillator system.
 It is defined by the Hamiltonian of two-dimensional oscillator, with the angular part replaced by a P\"oschl-Teller system on circle:
\be
\mathcal{H}_{TTW}=\frac{p^2_r}{2}+\frac{\mathcal{I}_{PT}}{r^2}+\frac{\omega^2r^2}{2},
\label{TTW}
\ee
\be
\mathcal{I}_{PT}=\frac{p^2_\varphi}{2}+\frac{k^2\alpha^2}{\sin^2k\varphi}+\frac{k^2\beta^2}{\sin^2k\varphi},
\label{PT}\ee
where $k$ is an integer.
It  coincides with the two-dimensional rational Calogero-oscillator model associated with the
dihedral group $D_{2k}$ \cite{lny}
and was initially considered as a new superintegrable model. The superintegrability was observed by  numerical simulations.
Later an  analytic expression for the
additional constant of motion was presented  \cite{kalnins}.

The two-dimensional Calogero-Coulomb system,  associated with dihedral group, is known as  a Post-Winternitz (PW) system.
It was  constructed
from the TTW system by performing the well-known Levi-Civita transformation,
 which maps the two-dimensional oscillator
into the Coulomb problem \cite{pw}. The PW system was also suggested as a new (independent from Calogero)
superintegrable  model.
It is also expressed via the P\"oschl-Teller Hamiltonian \eqref{PT},
\be
\mathcal{H}_{PW}=\frac{p^2_r}{2}+\frac{\mathcal{I}_{PT}}{r^2}-\frac{\gamma}{r}.
\label{PW}
\ee
In Ref.~\onlinecite{gonera1}, the superintegrability of the TTW-system  was explained from the  viewpoint of action-angle variable
formulation, while in
Ref.~\onlinecite{hlnsy},  using the same (action-angle) arguments,  the superintegrable generalizations of the TTW and PW systems   on sphere and
hyperboloid  were suggested. Below we briefly describe the constructions.

Consider an integrable $N$-dimensional system with the following Hamiltonian in action-angle variables:
\be
{\cal H}={\cal H}(nI_1+mI_2,I_3,\ldots, I_N),\qquad \{I_i,\Phi_j\}=\delta_{ij},\quad \Phi_i\in [0,2\pi),
\ee
where $n$ and $m$ are integers. The  Liouville integrals are expressed via the action variables $I_i$.
The system has a hidden symmetry, given by the additional constant of motion
\be
K_{hidden}={\rm Re}\; A(I_i){\rm e}^{\imath (m\Phi_1-n\Phi_2)},
\ee
where $A(I_i)$ is an arbitrary complex function on Liouville integrals.
Respectively, for the Hamiltonian
\be
{\cal H}={\cal H}(n_1I_1+n_2I_2+\ldots n_N I_N),
\ee
where $n_1,\ldots, n_N$ are integer numbers,
all  the functions
\be
K_{ij}={\rm Re}\; A_{ij}(I){\rm e}^{\imath(n_j\Phi_i-n_i\Phi_j)}.
\ee
are constants of motion, which are distinct  from the Liouville integrals.
The Liouville integrals  together with the additional integrals  $I_{i\,i+1}$ with $i=1,\ldots N-1$
constitute a set of $2N-1$  functionally independent constants of motion, ensuring the maximal superintegrability.

In Ref.~\onlinecite{hlnsy}  the integrable deformations of the   $N$-dimensional oscillator and  Coulomb systems
have been proposed on Euclidean space, sphere and  hyperboloid by replacing their
 angular part by an $(N-1)$-dimensional integrable system,
formulated in action-angle variables:
\be
H=\frac{p^{2}_r}{2}+\frac{{\cal I}(I_a)}{r^2}+V(r),
\qquad
\{p_r, r\}=1,
\qquad
\{I_a,\Phi^0_b\}=\delta_{ab},
\label{2}
\ee
where $ a,b=1,\ldots,N-1$ and
\be
 V_{osc}(r)=\frac{\omega^2r^2}{2}, \qquad V_{Coulomb}(r)=-\frac{\gamma}{r}.
\label{5}\ee
In other words, we obtain the deformation of the $N$-dimensional oscillator and Coulomb systems
by replacing the $SO(N)$ quadratic Casimir element $\mathbf{J}^2$, which defines the  kinetic part of the system on sphere
$\mathbb{S}^{N-1}$, with the Hamiltonian
 of some $(N-1)$-dimensional integrable system written in terms of the action-angle variables.

Next we have performed   similar analyses for the systems on $N$-dimensional sphere and (two-sheet) hyperboloid
with the oscillator and Coulomb potentials. These models were introduced, respectively, by Higgs \cite{higgs} and Schr\"odinger \cite{sch},
\be
\mathbb{S}^{N}:\quad H=\frac{p_{\chi}^2}{2 r_0^2}+\frac{{\cal I}}{ r_0^2 \sin^2\chi}+V(\tan\chi),
\quad \{p_\chi, \chi\}=1,
\label{3}\ee
\be
\mathbb{H}^{N}:\quad H=\frac{p_{\chi}^2}{2 r_0^2}+\frac{{\cal I}}{ r_0^2\sinh^2\chi}+V(\tanh\chi), \quad \{p_r, r\}=1
\label{4}
\ee
with  ${\cal I}$ depending on the (angular) action variables.
The exact forms for the potential are:
\begin{align}
\mathbb{S}^{N}:& \quad V_{Higgs}(\tan\chi)=\frac{r^2_0\omega^2\tan^2\chi}{2}, \qquad V_{Sch-Coulomb}(\tan\chi)=-\frac{\gamma}{r_0}\cot\chi,
\label{6}
\\
\mathbb{H}^{N}:& \quad V_{Higgs}(\tanh\chi)=\frac{r^2_0\omega^2\tanh^2\chi}{2}, \qquad V_{Sch-Coulomb}(\tanh\chi)=-\frac{\gamma}{r_0}\coth\chi.
\label{7}
\end{align}
 The following expressions for the Hamiltonians of oscillator-like systems had been derived:
\be
{\cal H}_{osc}={\cal H}_{osc}(2 I_r + \sqrt{2{\cal I}})=\left\{\begin{array}{ccc}
\omega (2 I_r + \sqrt{2{\cal I}})& {\rm for }& \mathbb{R}^N,\\
\frac{1}{2}(2I_\chi+ \sqrt{2{\cal I}}+\omega)^2- \frac{\omega^2}{2}&{\rm for}& \mathbb{S}^N,\\
-\frac{1}{2}(2I_\chi +\sqrt{2{\cal I}}-\omega)^2+ \frac{\omega^2}{2}&{\rm for}&  \mathbb{H}^N.
\end{array}
\right.
 \ee
 Respectively, the Hamiltonians of the  Coulomb-like systems read:
 \be
{\cal H}_{Coulomb}={\cal H}_{Coulomb}( I_r + \sqrt{2{\cal I}})=\left\{\begin{array}{ccc}
-\frac{\gamma^2}{2}(I_r + \sqrt{2{\cal I}})^2& {\rm for }& \mathbb{R}^N,\\
-\frac{\gamma^2}{2}(I_\chi + \sqrt{2{\cal I}})^2+\frac12(I_\chi+\sqrt{2{\cal I}})^2&{\rm for}& \mathbb{S}^N,\\
-\frac{\gamma ^2}{2} {(I_\chi-\sqrt{2 {\cal I}})^2} -\frac12(I_\chi-\sqrt{2 {\cal I}})^2&{\rm for}&  \mathbb{H}^N.
\end{array}
\right.
 \ee
Thus, it is easy to deduce that  for the angular Hamiltonian
\be
{\cal I}_{SphCalogero}=\frac 12\Big(\sum_{a=1}^{N-1} k_aI_a+\text{const}\Big)^2, \qquad k_a\in \mathbb{N},
\label{app}
\ee
the deformations of the oscillator and Coulomb systems become superintegrable.
In particular, the P\"oschl-Teller Hamiltonian has the same form \cite{lny}:
\be
\mathcal{I}_{PT}=\frac{k^2(I+\alpha+\beta)^2}{2}.
\ee
Hence, choosing $N=2$ and  $\mathcal{I}=\mathcal{I}_{PT}$,
we obtain the generalizations of the TTW and PW systems on sphere and hyperboloid
with additional constants of motion given by
 \be
\mathcal{K}_{TTW}={\rm Re}\; A(I){\rm e}^{\imath (k\Phi_r-2\Phi_\varphi)},\qquad \mathcal{K}_{PW}={\rm Re}\; A(I){\rm e}^{\imath (k\Phi_r-\Phi_\varphi)}.
 \ee
Here $\Phi_{\varphi}$ is the angle variable in the P\"oschl-Teller system,
and $\Phi_r$ is the angle variable  associated with $r$ and $p_r$. For explicit expressions,
see Ref.~\onlinecite{hlnsy}.

 Note that the angular part of the $N$-simensional rational Calogero model has  the form \eqref{app} as well.
 This is a reason for the superintegrabilty of the Calogero-oscillator and Calogero-Coulomb problems.
 It also suggests  that their superintegrable generalizations on the $N$-dimensional spheres and hyperboloids \cite{sphCal}.
Although the TTW and PW systems are  particular cases of the Calogero-type  models,
they continue to attract enough interest due to their simplicity.
In particular, a couple of years ago, Ranada suggested a specific  representation for the  constants of motion of the
TTW and PW systems (including those on sphere and hyperboloid) \cite{ranada},
called a "holomorphic factorization". For the TTW system it reads
\be
{\cal R}_{TTW}=(\bar M_0)^{k}N^2,
\label{ranadaTTWlint}
\ee
where
\be M_{0}=\frac{2 p_r}{r}\sqrt{2\cal{I}_{PT}}+2\imath \mathcal{H}_{TTW},
\label{ranadaTTW}
\ee
and
\be
 N=k(\beta-\alpha)+2\mathcal{I}_{PT}\cos{2k\varphi}+\imath\sqrt{2\mathcal{I}_{PT}}p_{\varphi}\sin 2k\varphi.
\label{ranadaN}\ee
A similar expression  exists in case of the (pseudo)spherical TTW system.
The additional constant of motion of PW system in Ranada's representation reads:
\be
\mathcal{M}_{PW}=({\bar M}_0)^{k}N,
\ee
and $N$ is given by Eq.~\eqref{ranadaN}, and
 \be
 M_0=p_r\sqrt{2\mathcal{I}_{PT}}+\imath\Big(\gamma-\frac{2\mathcal{I}_{PT}}{r}\Big) .
\ee
Such forms of the hidden constants of motion have a visible relation with their expressions in terms of the action-angle variables, which will be discussed below.
Hence,  the TTW and PW systems possess a natural  description  in spherical coordinates, where the "radial" part is separated
from the "angular" one. On the other hand, the radial parts are expressed via the generators of conformal algebra,
which can be  viewed as generators of isometries of the K\"ahler structure  of    Klein model of the Lobachevsky space \cite{lobach}.
Hence, we can represent phase spaces of the TTW and PW systems as a (semidirect) product of Lobachevsky space with cotangent bundle
of circle, and expect that the reformulation in these coordinates will help us to extend the expressions of hidden constants of motion to
higher dimensions. Similarly, phase spaces of the  $N$-dimensional oscillator and Coulomb systems and their Calogero-deformations
could be represented as a  semidirect product of Lobachevsky space and cotangent bundle on $(N-1)$-dimensional sphere \cite{sigma}.
One can expect, that Ranada's representation of hidden symmetries of the TTW and PW systems in these terms will take
a more transparent and elegant form. Furthermore, having in mind the relation of the TTW  and PW systems with rational Calogero models,
one can expect that the hidden symmetries of  latters could be represented in a similar way.

Hence, the purpose of this paper is to provide the planar TTW and PW systems and
their hidden constants of motion with such kind of formulation, as well as to discuss their extensions to higher dimensions.
A similar description of their (pseudo)spherical generalizations will be presented elsewhere.

\section{One-dimensional systems}
 Since the middle of seventies with Ref.~\onlinecite{fubini}  in the field-theoretical
literature
much attention has been paid  to a simple one-dimensional mechanical system
 given by the Hamiltonian
 \begin{equation}
   H_0 =\frac{{p}^2}{2}+\frac{g^2}{2x^2}.
\label{h}
\end{equation}
The reason was that it  forms
 the conformal algebra $so(1,2)$ \eqref{ca} together with the generators:
\begin{equation} D=px,\qquad K=\frac{x^2}{2} \label{dk}. \end{equation}

In Ref.~\onlinecite{lobach}  the following formulation of this is suggested.
Its   phase space   is parameterized by a single complex coordinate and identified with the Klein model
of the Lobachevsky plane:
\begin{equation} { z}=\frac{p}{x}+\frac{\imath g}{x^2}, \qquad {\rm
Im}\;{ z}>0: \qquad \{{ z},\bar{z}\}=-\frac{\imath}{g}\left({
z}-\bar {z} \right)^2. \label{px} \end{equation}

In this   parametrization,  the $so(1,2)$ generators (\ref{h}), (\ref{dk})
define the Killing potentials (Hamiltonian generators of the
isometries of the K\"ahler structure) of Klein model:
\begin{equation}
 H_0= g\frac{{ z} {\bar z}}{\imath({\bar z}-{ z})},
\qquad
D=g\frac{{ z}+\bar{ z}}{\imath ({\bar z}-{ z})},
\qquad
K= g\frac{1}{\imath ({ \bar z}- { z})}.
\label{uk}
\end{equation}
Let us remind, that the K\"ahler structure  is
\begin{equation} ds^2=-\frac{g d{ z} d\bar
{z}}{({\bar z}- { z})^2}. \label{klein}
\end{equation}
It is invariant under the
 discrete transformation
\begin{equation} { z}\to -\frac{1}{z}, \label{simt} \end{equation}
whereas the Killing
potentials (\ref{uk}) transform as follows:
 \begin{equation}
 H_0\to  K,\qquad K\to H_0,\qquad D\to -D.
\label{t}\end{equation}
Thus, it  maps $H_0$ to the free one-dimensional particle system. This
can be viewed as a one-dimensional analog of the decoupling transformation
of the Calogero Hamiltonian, considered in Refs.~\cite{decoupling}.

In order to construct a similar construction for higher-dimensional systems,
first, we introduce an appropriate "radial"
coordinate and  conjugated momentum, so that the higher-dimensional
system looks very similar to the one-dimensional conformal
mechanics. In that picture, the remaining "angular" degrees of
freedom are packed in the  Hamiltonian system on the
$(N-1)$-dimensional sphere, which replaces the coupling constant $g^2$
in the one-dimensional conformal mechanics. The angular Hamiltonian defines the constant
of motion of the is initial conformal mechanics.  Then
we relate the radial part of the $N$-dimensional conformal
mechanics with the Klein model of the Lobachevsky space, which is
completely similar to the aforementioned one-dimensional case.

\section{Higher-dimensional systems}
Let us consider the $N$-dimensional conformal mechanics, defined by
the  following Hamiltonian and symplectic structure:
\begin{equation}
\omega={d{\bf p} }\wedge {d{\bf x}},
\qquad
\mathcal{H}_0=\frac{{\bf p}^2}{2}+V({\bf x}),
\qquad
{\rm where}
\qquad ({\bf x}\cdot\nabla) V({\bf x} )=-2V({\bf x}).
\label{h1}
\end{equation}
 This Hamiltonian together with the generators
\begin{equation}
\mathcal{D}={\bf p}\cdot{\bf x},\qquad  \mathcal{K}=\frac{{\bf x}^2}{2} \label{dk1}
\end{equation}
 forms the
conformal algebra $so(1,2)$  (\ref{ca}). Here $\mathcal{D}$ defines the
dilatation and $\mathcal{K}$ defines  the conformal boost, ${\bf
x}=(x^1,\ldots, x^N)$, ${\bf p}=(p_1,\ldots, p_N)$.

Extracting the radius $r=|{\bf x}|$,  we can present the above
generators  in the following form:
\begin{equation} {\cal D}=p_r r ,
\qquad
{\cal K}=\frac{r^2}{2},
\qquad {\cal H}_0=\frac{{p}^2_r}{2}+
\frac{{\cal I}}{r^2},
\qquad {\cal I}\equiv \frac{{\bf J}^2}{2}+ U,
\qquad U\equiv r^2V({\bf r}).
\label{so2}
\end{equation}
 Here  $p_r=({\bf p}\cdot{\bf x})/r $ is the momentum,
conjugate to the radius:  $\{p_r, r\}=1$.
It is easy to check that $\mathcal{I}$ is the Casimir element  of conformal algebra $so(1.2)$:
\begin{equation}
4{\cal H}{\cal K}-{\cal D}^2= 2{\cal I}:
\qquad
\{\mathcal{I},\mathcal{H}_0\}=\{\mathcal{I},\mathcal{K}\}=\{\mathcal{I},\mathcal{D}\}=0.
\label{ml}
\end{equation}
Thus, it defines the constant of motion of the system \eqref{h1} and commutes with $r,p_r$ and,
hence, does not depend on them.
Instead, it depends on  spherical coordinates $\phi^a $ and
canonically conjugate momenta $\pi_a $.
As a Hamiltonian, ${\cal I}$ defines
the particle motion  on $(N-1)$-sphere in the potential $U(\phi^\alpha )$. The phase space is the
cotangent bundle  $T^*S^{N-1}$.
%

As in one dimension \cite{lobach} instead of the radial phase space
coordinates $r$ and $p_r$ we introduce  the following complex
variable (for simplicity, we restrict to $\mathcal{I}>1$):
\begin{equation} { z}=\frac{p_r}{r}+\frac{\imath \sqrt{2{\cal I}}}{r^2}\equiv
\frac{{\cal D}+\imath \sqrt {2{\cal I}}}{2{\cal K}},
\qquad
{\rm Im}\;{ z}>0.
 \label{px2}
\end{equation}
It obeys the following Poisson brackets:
\begin{equation}
\{{ z},\bar{ z}\}=-\frac{\imath}{\sqrt{2{\cal I}(u)}}\left({ z}-\bar { z}
\right)^2,
\label{k1}
\end{equation}
\begin{equation}
 \{u^\alpha,u^\beta\}=\omega^{\alpha\beta}(u),\qquad \{u^\alpha , z\}= (z -\bar z
)\frac{V^\alpha(u)}{2{\cal I}},\qquad\{u^\alpha, \bar z\}=( z -\bar z
)\frac{V^\alpha(u)}{2{\cal I}},\qquad
\end{equation}
where $ V^\alpha=\{u^\alpha, {\cal I}(u)\}$ are the
equations of motion of the angular system.

 The symplectic structure of the
conformal mechanics can be represented as follows:
 \begin{equation}
\Omega=-\imath \frac{\sqrt{2{\cal I}(u)} d{ z}\wedge  d\bar { z}}{({\bar z}-
{ z})^2} + \frac{(dz+d\bar z)\wedge d\sqrt{ 2{\cal I}(u)}}{\imath(\bar z- z
)}+ \frac{1}{2}\omega_{\alpha\beta}du^\alpha\wedge {du^\beta},
\end{equation}
  while the local
one-form, defining this symplectic structure, reads
 \begin{equation} \Omega=
d{\cal A},\qquad {\cal A}=\imath \sqrt{2\mathcal{I}(u)}\frac{dz+d\bar
z}{\imath(z-\bar z)} +A_0(u),\qquad dA_0=\frac{1}{2}
\omega_{\alpha\beta}du^\alpha\wedge du^\beta .
\end{equation}
 %
 Taking into account Eq.~(\ref{ml}), we can write:
\begin{equation}
\mathcal{H}_0=\sqrt{2{\cal I}(u)}\frac{{ z}\bar { z}}{\imath({ \bar z}- { z})},
\qquad \mathcal{D}= \sqrt{2{\cal I}(u)}\frac{{ z}+\bar { z}}{\imath ({\bar z}- {
z})}, \qquad \mathcal{K}= \frac{\sqrt{2{\cal I}(u)}}{\imath ({\bar z}- { z})},
\label{uk2}
\end{equation}
The  transformation (\ref{simt}) does not preserve the
symplectic structure, i. e., it is not a canonical transformation
for the  generic  conformal mechanics of dimension $d>1$.


Now we introduce the following generators, which will be used in our further considerations:
\be
M=\frac{z}{\sqrt{\imath({\bar z}- z )}}, \qquad \bar M=\frac{{\bar z}}{\sqrt{\imath(\bar z- z)}}.
\ee
 With the generators of the conformal algebra they form a highly nonlinear algebra:
\be
\{M,{\cal H}_0\}=\frac{\imath}{2}z\sqrt{\imath (\bar z -z)},
\qquad \{M,{\cal K}\}=\frac{2z}{\imath(\bar z-z)}, \qquad \{M,{\cal D}\}=\frac{z}{\sqrt{\imath(\bar z-z)}}=M,
\ee
\be
 \{M,\bar M\}= \frac{z-\bar z}{2\sqrt{2\mathcal{I}}}.
\ee

Let us introduce the angle-like  variable, conjugate with $\sqrt{2\mathcal{I}}$:
\be
\Lambda(u):
\qquad
 \big\{\Lambda,\sqrt{2\mathcal{I}}\big\}=1,
 \qquad \Lambda\in[0,2\pi).
\label{Phi}\ee
Using $M$ and $\Lambda$,  one can easily build a (complex) constant of motion for the conformal mechanics:
\be
{\cal M}=M{\rm e}^{\imath\Lambda},
\qquad \{{\cal M}, \mathcal{H}_0\}=0.
\ee
Evidently, its real part is the ratio of  Hamiltonian and its angular part
and does not contain any new constant of motion.
Nevertheless, such a complex representation seems to be useful not only from an aesthetical viewpoint, but also
for the construction of supersymmetric extensions.

Note that  we can write down the hidden symmetry generators for the conformal mechanics,
modified by the oscillator and Coulomb potentials  as well.
The Hamiltonian of the $N$-dimensional oscillator and its hidden symmetry generators  look
 as follows:
\be
\begin{gathered}
{\cal H}_{osc}={\cal H}_{0} +\omega^2{\cal K},
\quad {\cal M}_{osc}=\frac{z^2+\omega^2}{\imath(\bar z-z)}{\rm e}^{2\imath\Lambda}=\left(M^2+\omega^2{\cal K}\right){\rm e}^{2\imath\Lambda}\; :\quad
\{{\cal M}_{osc}, {\cal H}_{osc}\}=0
\end{gathered}
\ee
The Hamiltonian and  hidden symmetry of the Coulomb problem are defined by
 \be
 \begin{split}
 {\cal H}_{Coul}={\cal H}_0-\frac{\gamma}{\sqrt{2{\cal K}}},
  \quad{\cal M}_{Coul}=\left({{ M}}-\frac{\imath\gamma}{(8\sqrt{2\cal{I}})^{3/2}} \right){\rm e}^{\imath\Lambda}:\quad \{{\cal M}_{Coul}, {\cal H}_{Coul}\}=0,
 \end{split}
 \ee
The absolute values of both integrals do not produce anything new:
\be
|{\cal M}_{osc}|^2=\frac{\mathcal{H}^2_{osc}}{2\mathcal{I}}-\omega^2,
\qquad |{\cal M}_{Coul}|^2=\frac{\mathcal{H}_{Coul}}{\sqrt{2\mathcal{I}}}+\frac{\gamma^2}{2(\sqrt{2\mathcal{I}})^{3}}.
\ee
So, the hidden symmetry is encoded in their phase, depending on
the angular variables $\Phi(u)$.
Assume that the angular system
 is integrable.Hence the Hamiltonian
 and two-form are expressed in terms of the action-angle variables as follows:
 \[
\mathcal{I}=\mathcal{I}(I_a),
\qquad
\Omega=\sum_a dI_a\wedge d\Phi_a.
\]
Then the condition \eqref{Phi} implies the following local solutions   for $\Lambda$:
\be
\Lambda_a=\frac{\Phi_a}{\omega_a(I)}, \quad{\rm where}\quad \omega_a=\frac{\partial\sqrt{2\mathcal{I}}}{\partial I_a}.
\ee
Thus, to provide the global solution for a certain coordinate $a$,
we are forced to set $\omega_a(I)= k_a$ to a rational number:
\be
k_a=\frac{n_a}{m_a} ,\qquad m_a,n_a \in \mathcal{N}.
\ee

Then, taking $k_a$-th power for the locally defined conserved quantity, we get a globally defined
constant of motion for the system.
In this case, the hidden symmetry  of the  conformal mechanics reads:
\be
{\cal M}_a=M^{n_a}{\rm e}^{\imath m_a\Phi_a }.
\ee
Similarly, for the systems with oscillator and Coulomb potentials  one has:
\be
{\cal M}_{(a)osc}=\left(M^2+\omega^2{\cal K}\right)^{n_a}{\rm e}^{2\imath m_a\Phi_a},
\qquad
{\cal M}_{(a)Coul}=\left({{ M}}-\frac{\imath\gamma}{(8\sqrt{2\cal{I}})^{3/2}} \right)^{n_a}{\rm e}^{\imath m_a\Phi_a}.
\ee

To find the expression(s) for $\Phi$, let us remind
that the angular part  of these systems is just the quadratic Casimir element (angular momentum)
of $so(N)$ algebra on $(N-1)$- dimensional sphere, ${\cal I}=L_{N}^2/2$.
It can be decomposed by the eigenvalues of the embedded $SO(a)$ angular momenta $I_a $
as follows:
\be
{\cal I}=\frac12\left(\sum_{a=1}^{N-1}I_a \right)^2.
\ee

The functions $I_a$  and  their canonically conjugates $\Phi_a$
play the role of  the action-angle variables of the free particle on $S^{N-1}$.
{For details, see Appendix in Ref.~\onlinecite{hlnsy}.)
Their explicit forms are:
\be
I_a=\sqrt{j_a}-\sqrt{j_{a-1}},\quad {\rm where}\quad j_a=p^{2}_{a-1}+\frac{j_{a-1}}{\sin^2 \theta_{a-1}},\qquad a=1,\ldots N.
\ee
The related angle variables are:
\be
\Phi_a=\sum_{l=a}^{N-1}\arcsin(u_l)+\sum_{l=a+1}^{N-1}\arctan\Bigg(\sqrt{\frac{j_{l-l}}{j_l}}
\frac{u_l}{\sqrt{1-u_l^2}}\Bigg),
\ee
where
\be  u_a=\sqrt{\frac{j_a}{j_a-j_{a-l}}} \cos{\theta_a},
\qquad
\sqrt{j_a}=\sum_{m=1}^{a}I_m.
\ee
Hence, our expressions define
the $N-1$ functionally independent constants of motion
\be
{\cal M}_{(a)osc}=\left(M^2+\omega^2{\cal K}\right){\rm e}^{2\imath\Phi_a},
\qquad
{\cal M}_{(a)Coul}=\left({{ M}}+\imath\gamma \right){\rm e}^{\imath\Phi_a},
\ee
respectively,  for the $N$-dimensional oscillator and Coulomb problems.
Since these  systems have $N$ commuting constants of motion ($I_a$, ${\cal H}$), we
have obtained in this way the full set of their integrals.

To clarify the origin of these  generators, let us consider a particular case of two-dimensional systems.
The angular part   is a circle, and, respectively, $I=|p_\varphi|$, $\Phi=\varphi$ with $\varphi$
being a polar angle.
In this case, the oscillator Hamiltonian and its hidden constant of motion  read
\be
 H_{osc}=|p_\varphi|\frac{z\bar{z}+\omega^2}{\imath(\bar z-{z})},\qquad
\mathcal{M}_{osc}=\frac{i}{z-\bar{z}}(z^2+\omega^2)e^{2\imath\varphi}.
\ee
The latter can  also be presented as follows:
\be
\mathcal{M}_{osc}=\frac{H_1-H_2+2iH_{12}}{|p_{\varphi}|},\quad{\rm with} \quad H_{ab}= p_ap_b+\omega^2x_ax_b.
\ee
Here $H_{ab}$  is a standard  representation of the oscillator's hidden symmetry generators,
 sometimes called a Demkov tensor.

The Hamiltonian of two-dimensional Coulomb problem and its hidden symmetry generator are of the form

 \be
 H_{Coul}=|p_\varphi|\frac{z \bar{z}}{\imath(\bar z-{z})}-{\gamma}\sqrt{
 \frac{\imath(\bar z-{z}) }{2|p_\varphi |}},\qquad
{\cal M}_{Coul} = \Bigg(\frac{z}{\sqrt{\imath(\bar z-{z})}}-\frac{\imath\gamma}{\sqrt{2|p_\varphi|^3}}\Bigg)e^{\imath\varphi}
\ee
The Latter  is related with the components of the two-dimensional Runge-Lenz vector
${\bf A}=(A_x,A_y)$ as follows
\be
 \mathcal{M}_{Coul}=\frac{A_y-\imath A_x}{\sqrt{2|p_\varphi|^3}},\quad {\rm where}\quad
A_{x}=  p_\varphi p_{y}-\gamma \cos{\varphi},
\quad
A_{y}= p_\varphi p_{x}-\gamma \sin{\varphi}.
\ee

Now we are ready to apply this constructions to the TTW and PW systems.

\subsection{TTW and PW systems}
In order to formulate TTW and PW systems in the above terms,  we will use the action-angle formulation of the P\"oshl-Teller
Hamiltonian given in Ref.~\onlinecite{lny}:
\be
\mathcal{I}_{PT}=\frac{k^2{\tilde I}^2}{2}, \qquad {\tilde I}=I+\alpha+\beta,
\ee
where $I$ is an action variable.

The angle variable is related to the initial phase space coordinates as follows:
\be
a \sin(-2 \Phi) =\cos(2k\varphi)+b  ,
\qquad
 a=\sqrt{\Big(1-\frac{2(\alpha+\beta)}{{(k\tilde I)}^2}+b^2\Big)} ,
 \qquad
 b=\frac{\beta-\alpha}{(k{\tilde I})^2}.
\ee

Using the above expressions, we can present the Hamiltonian of TTW system and its hidden symmetry generator
 as follows:
\be
H_{TTW}=k\tilde{I}\frac{z \bar{z}+\omega^2}{\imath(\bar z-{z})} ,
\qquad
{\cal M}_{TTW}=\Big(\frac{z^2+\omega^2}{\imath(\bar{z}-z)}\Big)^k e^{2\imath \Phi}.
\ee
The Ranada's constant of motion is related with the above one:
\be
K= -a^2 \frac{(2k\tilde{I})^{2k+4}}{16}\Big(\frac{\bar{z}^2+\omega^2}{z-\bar{z}}\Big)^{2k}e^{-4\imath \Phi}
=-a^2 \frac{(2k\tilde{I})^{2k+4}}{16}{\cal \bar M}_{TTW}^2.
\ee

 We repeat the same procedure   for the PW system as well.
Using the expressions  for action-angle variables of the P\"oschl-Teller Hamiltonian, we get:
 \be
 \mathcal{H}_{PW}=ik\tilde{I}\frac{\bar{z}z}{z-\bar{z}}-\frac{\gamma}{2k\tilde{I}}\sqrt{i(\bar{z}-z)} ,
 \qquad
  {\cal M}_{PW}=\left(\frac{z}{\sqrt{i(\bar{z}-z)}}-\frac{i\gamma}{k\tilde{I}\sqrt{2k\tilde{I}}}\right)^{k}e^{i\Phi}.
 \ee
Respectively, the Ranada's constant of motion takes the form
\be
 K=-ia(k\tilde{I})^2\left(k\tilde{I}\sqrt{2k\tilde{I}}\frac{z}{\sqrt{i(\bar{z}-z)}}+{i\gamma}\right)^{2k} e^{2i\Phi}=
 -ia(k\tilde{I})^{2k+2}{\cal \bar M}^2_{PW}.
 \ee

\acknowledgments
The authors thank Vahagn Yeghikyan for useful comments.
This work was partially supported
by the Armenian State Committee of Science grants 15RF-039  and 15T-1C367.
It was done within programs of  ICTP Network NET68 and of the Regional Training Network on Theoretical Physics
sponsored by Volkswagenstiftung Contract no. 86 260.

\end{document}